\documentclass[twoside,12pt]{article}
\usepackage{CJK}
\usepackage{indentfirst,amsmath,amssymb,hyperref}
\usepackage{bm}
\usepackage{epsfig}

\def\figwidth{0.75\textwidth}

\def\mnras{MNRAS}
\def\apj{Astrophys. J.}
\def\apjl{Astrophys. J. Lett.}

\def\prl{Phys. Rev. Lett.}
\def\prd{Phys. Rev. D}
\def\apss{Astrophysics and Space Science}
\def\jcap{JCAP}

\begin{document}

\begin{CJK*}{GBK}{song}

\thispagestyle{empty} \vspace*{0.8cm}\hbox
to\textwidth{\vbox{\hfill\huge\sf Commun. Theor. Phys.\hfill}}
\par\noindent\rule[3mm]{\textwidth}{0.2pt}\hspace*{-\textwidth}\noindent
\rule[2.5mm]{\textwidth}{0.2pt}


\begin{center}
\LARGE\bf Thawing k-essence dark energy in the PAge space 
\end{center}

\footnotetext{\hspace*{-.45cm}\footnotesize $^\dag$Corresponding author, E-mail: huangzhq25@mail.sysu.edu.cn  }

\begin{center}
  Zhiqi Huang$^{\rm a, b)\dagger}$ 
\end{center}

\begin{center}
\begin{footnotesize} \sl
${}^{\rm a)}$ School of Physics and Astronomy, Sun Yat-sen University, 2 Daxue Road, Tangjia, Zhuhai, 519082, China \\
${}^{\rm b)}$ CSST Science Center for the Guangdong-Hongkong-Macau Greater Bay Area, Sun Yat-sen University, 2 Daxue Road, Tangjia, Zhuhai, 519082, China 
\end{footnotesize}
\end{center}

\begin{center}
\footnotesize (Received XXXX; revised manuscript received XXXX)

\end{center}

\vspace*{2mm}

\begin{center}
  \begin{minipage}{15.5cm}
  \parindent 20pt\footnotesize

  A broad class of dark energy models can be written in the form of k-essence, whose Lagrangian density is a two-variable function of a scalar field $\phi$ and its kinetic energy $X\equiv \frac{1}{2}\partial^\mu\phi \partial_\mu\phi$. In the thawing scenario, the scalar field becomes dynamic only when the Hubble friction drops below its mass scale in the late universe. Thawing k-essence dark energy models can be randomly sampled by generating the Taylor expansion coefficients of its Lagrangian density from random matrices~\cite{thaws}. Ref.~\cite{thaws} points out that the non-uniform distribution of effective equation of state parameters $(w_0, w_a)$ of thawing k-essence model can be used to improve the statistics of model selection. The present work studies the statistics of thawing k-essence in a more general framework that is Parameterized by the Age of the universe (PAge)~\cite{PAge}. For fixed matter fraction $\Omega_m$, the random thawing k-essence models cluster in a narrow band in the PAge parameter space, providing a strong theoretical prior. We simulate cosmic shear power spectrum data for the Chinese Space Station Telescope optical survey, and compare the fisher forecast with and without the theoretical prior of thawing k-essence. For an optimal tomography binning scheme, the theoretical prior improves the figure of merit in PAge space by a factor of $3.3$.
  
\end{minipage}
\end{center}

\begin{center}
\begin{minipage}{15.5cm}
\begin{minipage}[t]{2.3cm}{\bf Keywords:}\end{minipage}
\begin{minipage}[t]{13.1cm}
dark energy, cosmological parameters, large-scale structure of Universe
\end{minipage}\par\vglue8pt

\end{minipage}
\end{center}

\section{Introduction}

Since the discovery of the accelerated expansion of the late Universe~\cite{Riess98, Schmidt98, Perlmutter99}, it has been widely accepted that the current Universe is dominated by a dark energy component with negative pressure, whose microscopic nature is often interpreted as a cosmological constant (vacuum energy) that is conventionally denoted as $\Lambda$. Over the past two decades, the standard six-parameter $\Lambda$ cold dark matter ($\Lambda$CDM) model has been confronted with a host of observational tests. The high-precision measurements of the temperature and polarization anisotropies of the cosmic microwave background (CMB) provide so far the most stringent constraints on the cosmological parameters~\cite{WMAP9yr, Planck2018}, which agree well with many other observations such as the baryon acoustic oscillations (BAO)~\cite{6dF:BAO, BOSS16, BOSS:Lya, BOSS:quasar, eBOSS:BAO}, the Type Ia supernovae (SN)~\cite{JLA, Pantheon}, the redshift-space distortion (RSD)~\cite{eBOSS:RSD, VIPERS:RSD}, and the cosmic chronometers (CC)~\cite{CC01,Simon:2004tf,Stern:2009ep,Zhang:2012mp,Moresco:2012jh,Moresco:2015cya,Moresco:2016mzx,Ratsimbazafy:2017vga}.

Despite the observational success, the extraordinary smallness of the vacuum energy (fine-tuning problem) and the coincidence that $\Lambda$ dominates the universe only recently (coincidence problem) have, at least philosophically, disturbed cosmologists for decades~\cite{Weinberg89}. Moreover, as the accuracy of observations improves, the great observational success of the $\Lambda$CDM model is now challenged by a few anomalies. For instance, the Hubble constant $H_0$ inferred from CMB + $\Lambda$CDM is in $\sim 5\sigma$ tension with the distance-ladder measurements~\cite{Riess21, Riess22}. Less significant challenges include a $3.4\sigma$ tension in the matter density fluctuation parameter $S_8$ between CMB and some cosmic shear data~\cite{KiDS20, KiDS22, Amon_etal}, a $2.8\sigma$ excess of lensing smearing in the CMB power spectra~\cite{Planck2018}, and the lack of large-angle correlation in CMB temperature~\cite{COBE96,WMAP03,Efstathiou10, Planck2015IandS}, etc. See Ref.~\cite{Challenges} for a recent comprehensive review of the observational challenges to the $\Lambda$CDM model.

Given that $\Lambda$ might not be the ultimate truth, we are well motivated to construct alternative dark energy models. A simple and in some sense also minimal construction is to introduce a scalar degree of freedom. Because high-order derivative theories typically suffer from the Ostrogradsky instability~\cite{Ostrogradsky}, it is often assumed that the Lagrangian density only depends on the scalar field value and its kinetic energy $X = \frac{1}{2}\partial_\mu\phi\partial^\mu\phi$. This class of dark energy models, often dubbed as k-essence models, allows a variety of cosmological solutions with rich phenomena~\cite{Chiba00, Malquarti03, Chimento04, Aguirregabiria04, Chimento05, Kim05, Lazkoz05, Aguirregabiria05, Chimento06, Rendall06, Cross, Putter07, Cruz09, Gao10, Chimento10, DGP10, Chimento11, Deffayet11, Tsyba11, fT12a, fT12b, De-Santiago12, Sharif12, RF12, Cardenas15, Graham15, BMM16, Horndeski16, Tannukij16, UnifiedDMDE16, Cordero19, Mukherjee21, Shi21, Barvinsky21, Tian21, OQ21, Lara22}. In the early time when k-essence dark energy was first proposed, interests were more focused on using the so-called tracking solutions, where the field has attractor-like dynamics in the early universe, to resolve the coincidence problem~\cite{Kessence_APC, Chiba02, Essentials, Das06}. It was understood later that the tracking k-essence models are not very successful solutions to the coincidence problem, because they require additional fine-tuning and superluminal fluctuations~\cite{Coincidence, NoGo, NoNoGo}. Moreover, tracking models typically predict moderate deviation from $\Lambda$, which is more and more disfavored as the accuracy of observations improves~\cite{Planck2015DE, Planck2018}. Alternatively, one can consider the so-called thawing k-essence~\cite{Scherrer2008, Chiba2009, WZ_Chiba, WZ_Kehayias, thaws}, whose mass scale is close to or less than the current expansion rate of the Universe. In the thawing picture, the k-essence field is frozen by the large Hubble friction in the early Universe. Only at low redshift when the expansion rate drops below its mass scale, the field starts to roll. The lightness assumption (mass $\lesssim H_0$) of thawing k-essence naturally leads to non-clustering dark energy whose perturbations are suppressed on sub-horizon scales. There do exist, however, models of dark energy with noticeable sub-horizon perturbations~\cite{Feng09, Bueno10, Gubitosi13, RunMp, Bellini18, Creminelli20}. In the present work we do not discuss clustering dark energy models, as they typically need to be treated in a one-by-one manner.

The assumption of thawing scenario significantly reduces the model complexity. By generating the Taylor expansion coefficients of $\mathcal{L}(\phi, X)$ from random matrices, Ref.~\cite{thaws} shows that a majority of k-essence dark energy models follows an approximate consistency relation $w_a\approx  -1.42 \left(\frac{\Omega_m}{0.3}\right)^{0.64}(1+w_0)$, where $\Omega_m$ is the present matter density fraction and $w_0, w_a$ are the Chevallier-Polarski-Linder (CPL) parameters for dark energy equation of state~\cite{Chevallier:2000qy,Linder:2002et}. The consistency relation can be understood as follows. Due to the thawing nature, the present rolling speed of the scalar field, which is characterized by $1+w_0$, is typically correlated to the acceleration of late-time rolling, which is characterized by $w_a$.


The approximate consistency relation can be combined with observational data to improve the constraining power of cosmological data, which is often measured with the so-called figure of merit in marginalized $w_0$-$w_a$ space. For a concrete model, however, the dark energy equation of state does not exactly follow the CPL form $w(a) = w_0+w_a(1-a)$, where $a$ denotes the scale factor. The parameters $w_0, w_a$ therefore only have an approximate meaning and should be considered as an effective description of dark energy at low redshift. In the present work, we consider another effective description of dark energy with the Parameterization based on cosmic Age (PAge)~\cite{PAge, PAgeSN, PAgeGRB, PAgeRSD, MAPAge, Cai22a, Cai22b}. Compared to the CPL $w_0$-$w_a$ effective description, PAge does not suffer from a strong parameter degeneracy that is commonly found between $w_0$ and $w_a$. Thus, the parameter space of PAge is more compact. The FiguRe Of Merit for the parameterization based on cosmic Age, which we abbreviate as FROMAge to show our French taste, is an equally good, if not better, indicator of the constraining power of cosmological data.

The article is organized as follows. Section~2 briefly reviews PAge cosmology. In section~3, we use the numerical tool developed in Ref.~\cite{thaws} to generate an ensemble of random thawing k-essence dark energy models, which are then mapped into PAge parameter space. In section~4, we  take a future cosmic shear survey as a working example to quantify by how much the  thawing k-essence prior may improve the constraining power of cosmological data. Section~5 concludes. Throughout the paper we work with natural units $c=\hbar=1$ and a spatially-flat Universe with Friedmann-Lema\^itre-Robertson-Walker background. The cosmological time and Hubble parameter are denoted as $t$ and $H$, respectively. The dark energy equation of state (EOS) is denoted as $w$, which in general is a function of redshift $z$. A dot represents derivative with respect to the cosmological time. The current scale factor is normalized to unity. The Hubble constant is denoted as $H_0=100 h\,\mathrm{km/s/Mpc}$. The square root of the cosmic variance of the mean density in a sphere with radius $8h^{-1}\mathrm{Mpc}$ is denoted as $\sigma_8$, which then defines the  $S_8\equiv \sigma_8\left(\frac{\Omega_m}{0.3}\right)^{0.5}$ parameter.

\section{PAge cosmology \label{sec:page}}

At redshift $z\lesssim 100$, where the radiation component can be ignored, PAge approximates the expansion history of the Universe with the following ansatz~\cite{PAge}
\begin{equation}
  \frac{H}{H_0} = 1 + \frac{2}{3}\left(1-\eta\frac{H_0t}{p_{\rm age}}\right)\left(\frac{1}{H_0t} - \frac{1}{p_{\rm age}}\right), \label{eq:page}
\end{equation}
where $p_{\rm age} = H_0t_0$ is the age of the Universe measured in unit of $H_0^{-1}$ and $\eta<1$ is a phenomenological parameter approximately describing the deviation from an Einstein de-Sitter universe.

Although it may seem like a casual assumption, the PAge ansatz~\eqref{eq:page} makes use of quite a few physical conditions. First of all, the parameters $H_0$ and $p_{\rm age}$ are physical quantities that can be directly computed for any given physical model. Secondly, ansatz~\eqref{eq:page} automatically sets the matter-dominated behavior at high redshift ($\lim_{t\rightarrow 0^+}Ht = \frac{2}{3}$). Finally, ansatz~\eqref{eq:page} guarantees that the expansion rate $H$ monotonically decreases as the Universe expands. Thanks to these physically motivated features, PAge well approximates many dark energy and modified gravity models~\cite{PAge, PAgeSN}, and performs better than many other phenomenological approaches, such as the oft-used polynomial approximation~\cite{Jerk}.

At the background level, when $H_0^{-1}$ is treated as a time unit, the expansion history is determined by $p_{\rm page}$ and $\eta$, and therefore $\Omega_m$ is not a parameter in PAge. While perturbation calculation is needed for the simulation of the cosmic shear data, we add $\Omega_m$ to the PAge framework and employ the following linear growth equation
\begin{equation}
   \frac{d^{2}D }{dt^{2}}+2H\frac{dD }{dt}-\frac{3H_{0}^{2}}{2a^{3}}\Omega _{m}D =0. \label{eq:growth}
 \end{equation}
The assumption goes into the above equation is that dark energy perturbations at sub-horizon and linear scales can be ignored.

Although more sophisticated approaches exist, for simplicity and to show the robustness of PAge approximation, we follow the simple method in Ref.~\cite{PAge} to map dark energy models to PAge space. The $\eta$ parameter is calculated using the deceleration parameter $q\equiv - \frac{a\ddot a}{\dot a^2}$ evaluated at redshift zero.

\section{Thawing k-essence in PAge space \label{sec:model}}

We use the numerical tool developed in Ref.~\cite{thaws}, which has been made publicly available at \url{http://zhiqihuang.top/codes/scan_kessence.tar.gz}, to generate random k-essence dark energy models. The program settings are shown in Table~\ref{tab:settings}. See also Ref.~\cite{thaws} for more detailed documentation of the program parameters.

\begin{table}
  \caption{k-essence generator program settings \label{tab:settings}}
  \begin{tabular}{llll}
    \hline
    \hline
    term in~\cite{thaws} & program variable & definition &  value \\
    \hline
    $n$ & dimX, dimV &  Taylor expansion truncation &  $10$ \\
    $\sigma$ & rand\_width & sampling width & 3 \\
    $k_{\rm pivot}$ & khMpc\_pivot & wavenumber to compute perturbations & 0.05 \\
    ultraviolet stability   & min\_cs2 & lower bound for sound speed & $0$ \\
    acceleration   & max\_w &  upper bound for dark energy EOS & $-\frac{1}{3}$ \\
    smoothness   & max\_growth & upper bound for growth of perturbations & $100$ \\
    thawing condition   & frozen\_cut & upper bound for early-Universe $|1+w|$ & $0.01$ \\    
    \hline
  \end{tabular}
\end{table}

It has been shown in Ref.~\cite{thaws}, and also tested in the present work, that increasing the truncation order and the sampling width do not change much the distribution of sampled trajectories. This is because models with increasing complexity typically violate the thawing condition ($|1+w|\ll 1$ in the early Universe), the acceleration assumption ($w<-\frac{1}{3}$) or the smoothness assumption (growth of density contrast $\lesssim 10^2$), and thus are rejected by the program.

\begin{figure}
\begin{center}  
\includegraphics[width=0.48\textwidth]{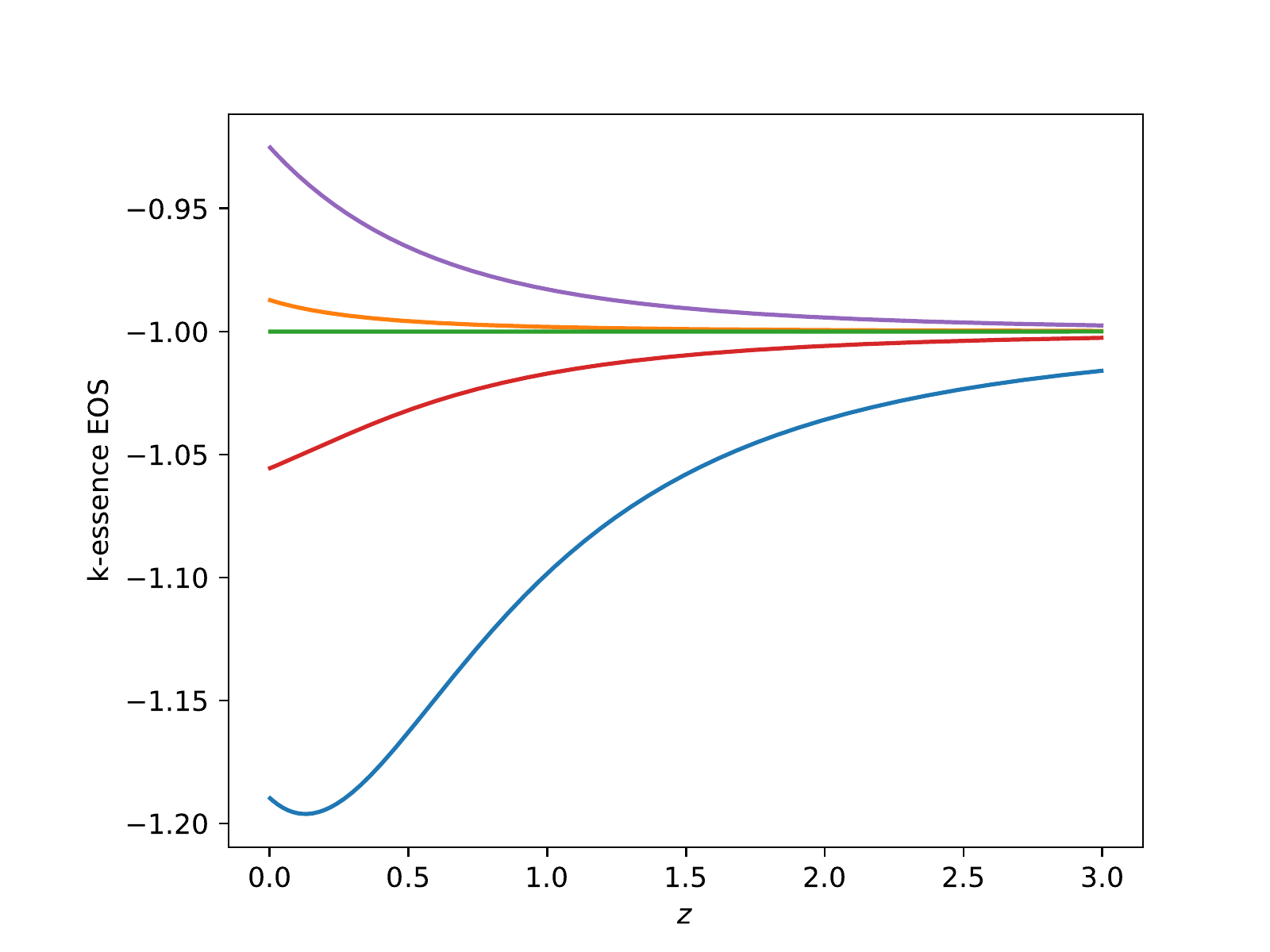}\includegraphics[width=0.48\textwidth]{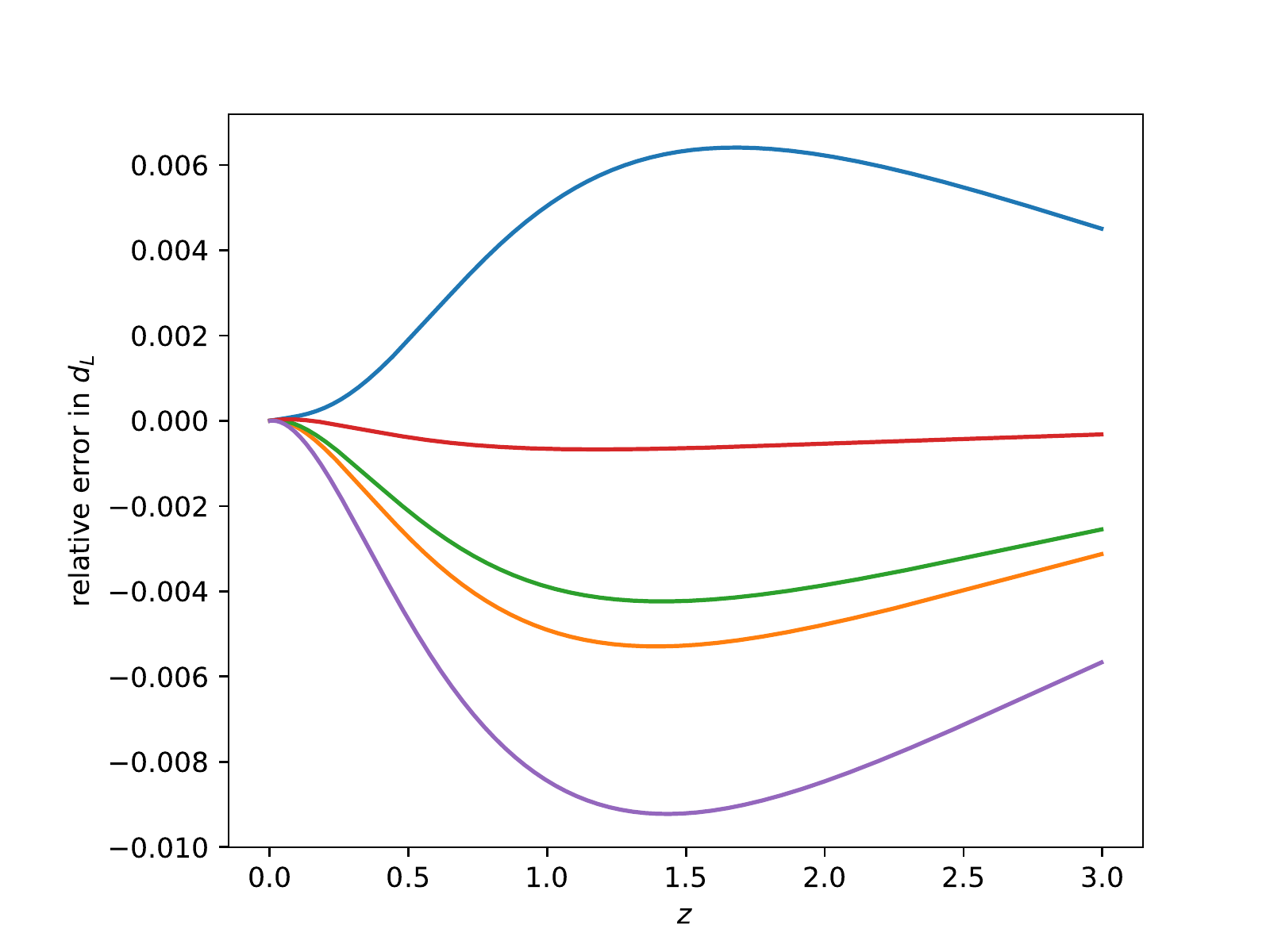}
\caption{The accuracy of PAge approximation. Left panel: EOS $w(z)$ of a few randomly sampled k-essence dark energy models; Right panel: relative error in luminosity distances when the models in the left panel are approximated with PAge. In all cases $\Omega_m$ is fixed to $0.3$.\label{fig:fit}}
\end{center}
\end{figure}

We generate 41000 random k-essence dark energy models for a flat prior $\Omega_m\in [0.25, 0.35]$. The models are then mapped into PAge space to generate a joint distribution of $(p_{\rm page}, \eta, \Omega_m)$, which we refer to as the thawing k-essence prior. The mapping procedure comes with a tiny accuracy loss in predictions of cosmological observables. In the left panel of Figure~\ref{fig:fit}, we show a few k-essence dark energy EOS trajectories with different colors. The relative difference between the luminosity distances predicted from each model and that from its PAge approximation is shown with the same color in the right panel. The errors are typically at sub-percent level. These tiny errors may be relevant for future cosmological surveys and can be corrected with a more sophisticated approach~\cite{MAPAge}. We nevertheless work in the original simple PAge framework that is easier to interpret, because the main purpose of the present work is to study the impact of the thawing k-essence prior, rather than the accuracy of PAge approximation.

\begin{figure}
\begin{center}  
\includegraphics[width=\figwidth]{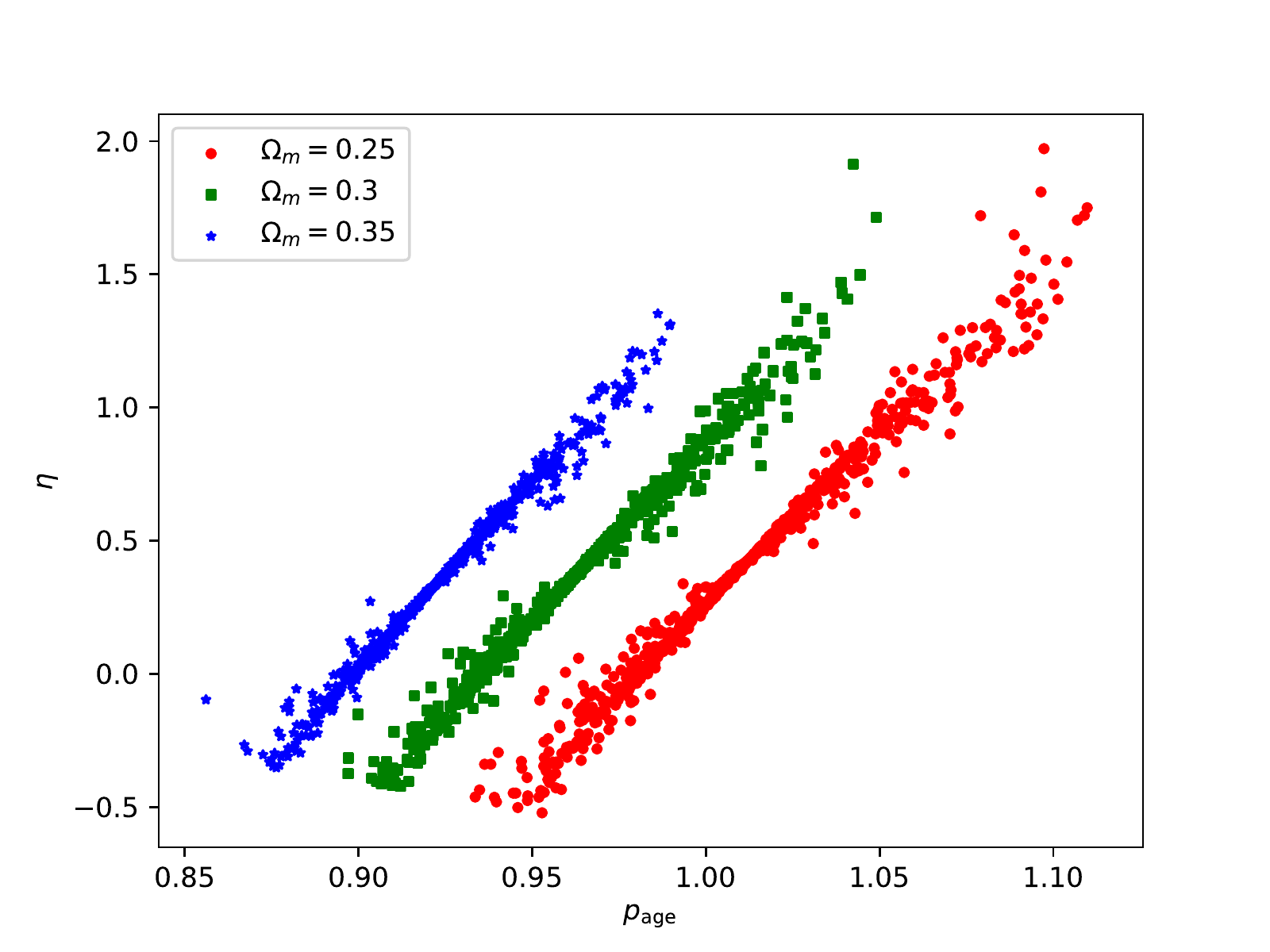}
\caption{Randomly sampled k-essence dark energy models mapped into the PAge space.\label{fig:scatter}}
\end{center}
\end{figure}

Due to parameter degeneracy, if the dark energy EOS is a free function of redshift, exact reconstruction of $\Omega_m$ from the expansion history of the universe is impossible. Since the Lagrangian density $\mathcal{L}(\phi, X)$ is a free function, the EOS of k-essence is almost free, too. However, when the aforementioned physical assumptions are applied, the EOS of thawing k-essence dark energy is not free in a statistical sense. In Figure~\ref{fig:scatter} we compare the mapped $(p_{\rm age}, \eta)$ samples for $\Omega_m = 0.25$, $0.3$ and $0.35$, respectively. It is evident that one can obtain a statistical constraint on $\Omega_m$ from the evolution history that is determined by $(p_{\rm age}, \eta)$. This is a non-trivial result. For a cosmic shear survey, the additional information on $\Omega_m$ can break the strong degeneracy between $\Omega_m$ and $\sigma_8$ and lead to a better reconstruction of low-redshift physics. To make the idea more concrete, in the next section we take a future cosmic shear survey as a working example to quantify the impact of the thawing k-essence prior.

\section{Cosmic Shear Fisher Forecast \label{sec:fisher}}

To make the analysis simple and easy to interpret, we only consider the statistics of the convergence field. The angular power spectrum between the redshift bins $i$ and $j$ is given by the Limber approximation~\cite{Limber54, Kaiser92, Kaiser98, Hu99}
\begin{equation}
  C_{\ell; i, j} = \int_0^\infty  W_i(z) W_j(z) P_m\left(k=\frac{\ell}{\chi(z)};z\right)\, \frac{d\chi}{dz}\, dz,
\end{equation}
where the comoving angular diameter distance in a spatially flat universe is given by
\begin{equation}
  \chi(z) = \int_0^z \frac{d z'}{H(z')}.
\end{equation}
 The non-linear matter power spectrum at redshift $z$, $P_m(k; z)$ where $k$ denotes the wavenumber, is calculated with the Bardeen-Bond-Kaiser-Szalay (BBKS) fitting formula~\cite{BBKS} and the halo-fit formula~\cite{Smith03,Takahashi12}. The weight function in the $i$-th bin $z\in\left[z_i^{\min}, z_i^{\max}\right]$ is given by
\begin{equation}
  W_i(z) = \left\{\begin{array}{ll}
                    \frac{3}{2}\Omega_mH_0^2 (1+z) \frac{1}{\bar{n}_i}\int_{ \max(z, z_i^{\min})}^{z_i^{\max}} \left(1-\frac{\chi(z)}{\chi(z')}\right) \frac{dn}{dz'} dz', & \ \text{ if } z < z_i^{\max}, \\
                    0, & \ \text{ if } z\ge z_i^{\max},
                  \end{array}
                  \right.
\end{equation}
where $\frac{dn}{dz}$ is the observed galaxy number per unit sky area per unit redshift. The observed galaxy number density in the $i$-th bin is an integral
\begin{equation}
  \bar{n}_i = \int_{z_i^{\min}}^{z_i^{\max}} \frac{dn}{dz} dz.
\end{equation}
The total number density of observed galaxies is then the sum $n_{\rm total} = \sum_i n_i$.

The observed convergence power spectrum with shot noise is modeled as
\begin{equation}
  C_{\ell; i, j}^{\rm obs} = C_\ell + \frac{\sigma_{\epsilon}^2}{\bar{n}_i}\delta_{ij}, \label{eq:Clobs}
\end{equation}
where $\delta_{ij}$ is the Kronecker delta function and $\sigma_\epsilon$ is the root mean square of the galaxy intrinsic ellipticity.

For the angular scales we take a conservative multipole range $10\le \ell \le 2500$. Due to the central limit theorem, the integrated convergence fields over this range are quite close to Gaussian~\cite{Scoccimarro99, White00, Cooray01}, and therefore can be written as 
\begin{equation}
  \mathrm{Cov}\left[C_{\ell_1; i_1, j_1}^{\rm obs}, C_{\ell_2; i_2, j_2}^{\rm obs}\right] =
  \frac{\delta_{\ell_1\ell_2}}{\left(2\ell_1+1\right)f_{\rm sky}}\left(C_{\ell_1; i_1, i_2}^{\rm obs} C_{\ell_2; j_1, j_2}^{\rm obs} + C_{\ell_1; i_1, j_2}^{\rm obs} C_{\ell_2; i_2, j_1}^{\rm obs} \right), \label{eq:cov}
\end{equation}
where $f_{\rm sky}$ is the fraction of sky that is observed. 

\begin{figure}
  \begin{center}  
    \includegraphics[width=\figwidth]{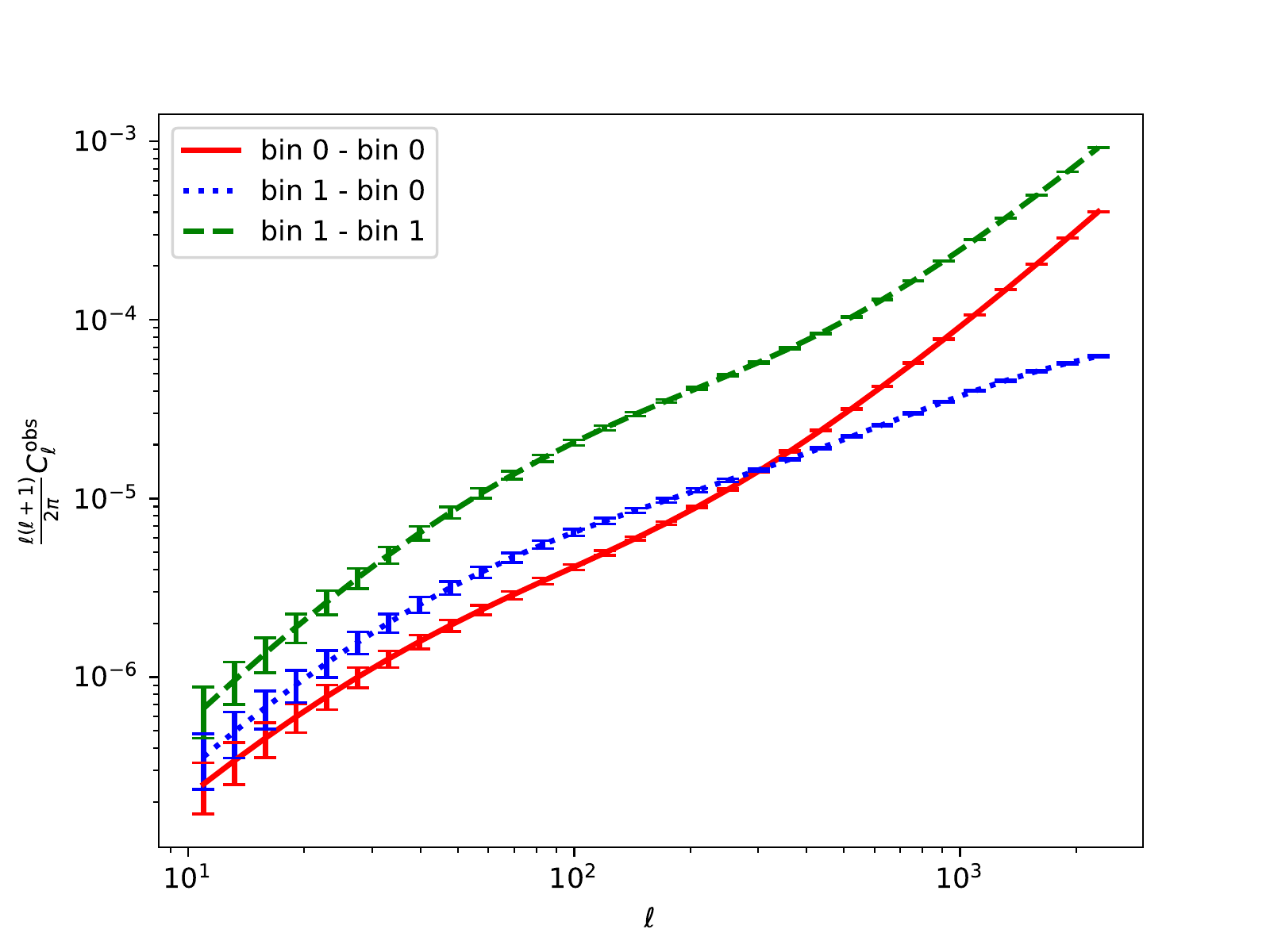}
    \caption{Simulated cosmic shear data with two redshift bins: $z\in [0, 1]$ (bin 0) and $z\in [1, 3]$ (bin 1).\label{fig:sim}}
  \end{center}
\end{figure}

If the cosmological redshifts of galaxies were all perfectly known, an optimal analysis would be done in the limit of taking infinitely many redshift bins. In practice, however, the redshift of a photometric survey has a large uncertainty, which in our simulation is assumed to be $\sigma(z)=0.03(1+z)$. Conventionally when doing Fisher forecast, the photo-z errors are treated by marginalizing some shift parameters and spreading parameters~\cite{Hu99, Ma06}, and the result inevitably depends on many assumptions that are difficult to justify at the stage of forecasting. To make the result robust and easy to interpret, we take a very conservative approach by simply discarding the galaxies samples around the edges of the redshift bins. More concretely, we cut each redshift bin $\left[z_i^{\min}, z_i^{\max}\right]$ to a smaller one $\left[z_i^{\min}+\sigma(z_i^{\min}), z_i^{\max} - \sigma(z_i^{\min})\right]$. This approach is conservative because we have assumed almost no knowledge about the photo-z error distribution function, which in realistic surveys will be known to some extent.

We have assumed that many other subtle effects such as the intrinsic alignment contamination~\cite{Takada04b}, catastrophic redshift outliers~\cite{Huterer06}, and the super-sample covariance~\cite{Takada13} can be well calibrated. The reader is referred to Refs.~\cite{Huterer06, Li14, Takada14, Takahashi14, Li14b, Kilbinger15} for more detailed discussion about calibration of these systematics.

In our simulation we assume a galaxy intrinsic ellipticity $\sigma_\epsilon=0.3$,  a galaxy distribution $n(z) \propto z^2e^{-z/0.3}$ that is normalized by $n_{\rm total} = 28\,\mathrm{arcmin}^{-2}$, and a sky coverage $f_{\rm sky} = 0.424$. The configuration roughly corresponds to the optical survey that will be carried out by the Chinese Space Station Telescope~\cite{CSST}. In Figure~\ref{fig:sim}, we show the simulated $C_\ell^{\rm obs}$ and their standard deviations for two redshift bins and thirty $\ell$-bins.

\begin{figure}
\begin{center}  
  \includegraphics[width=0.48\textwidth]{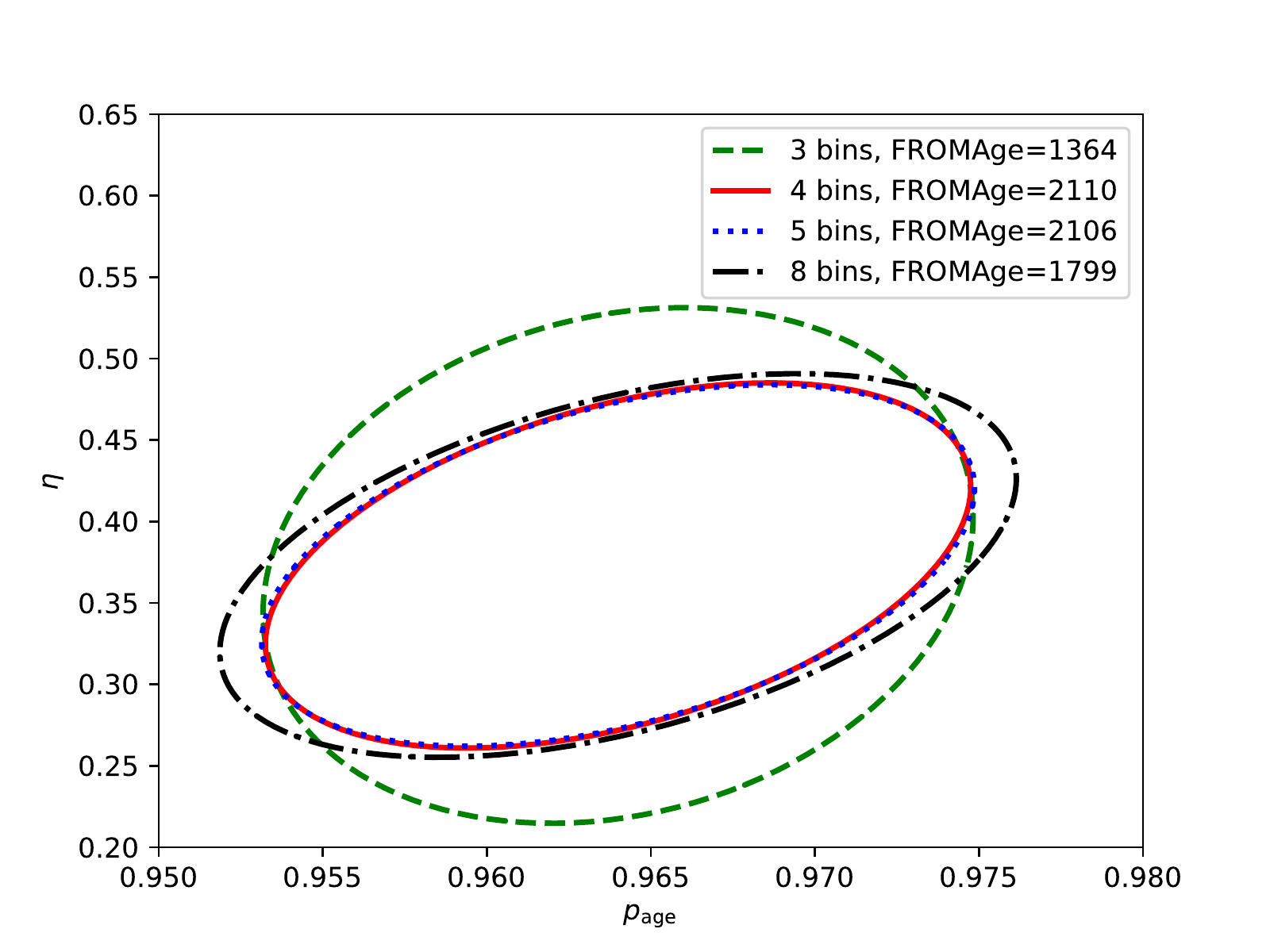}\includegraphics[width=0.48\textwidth]{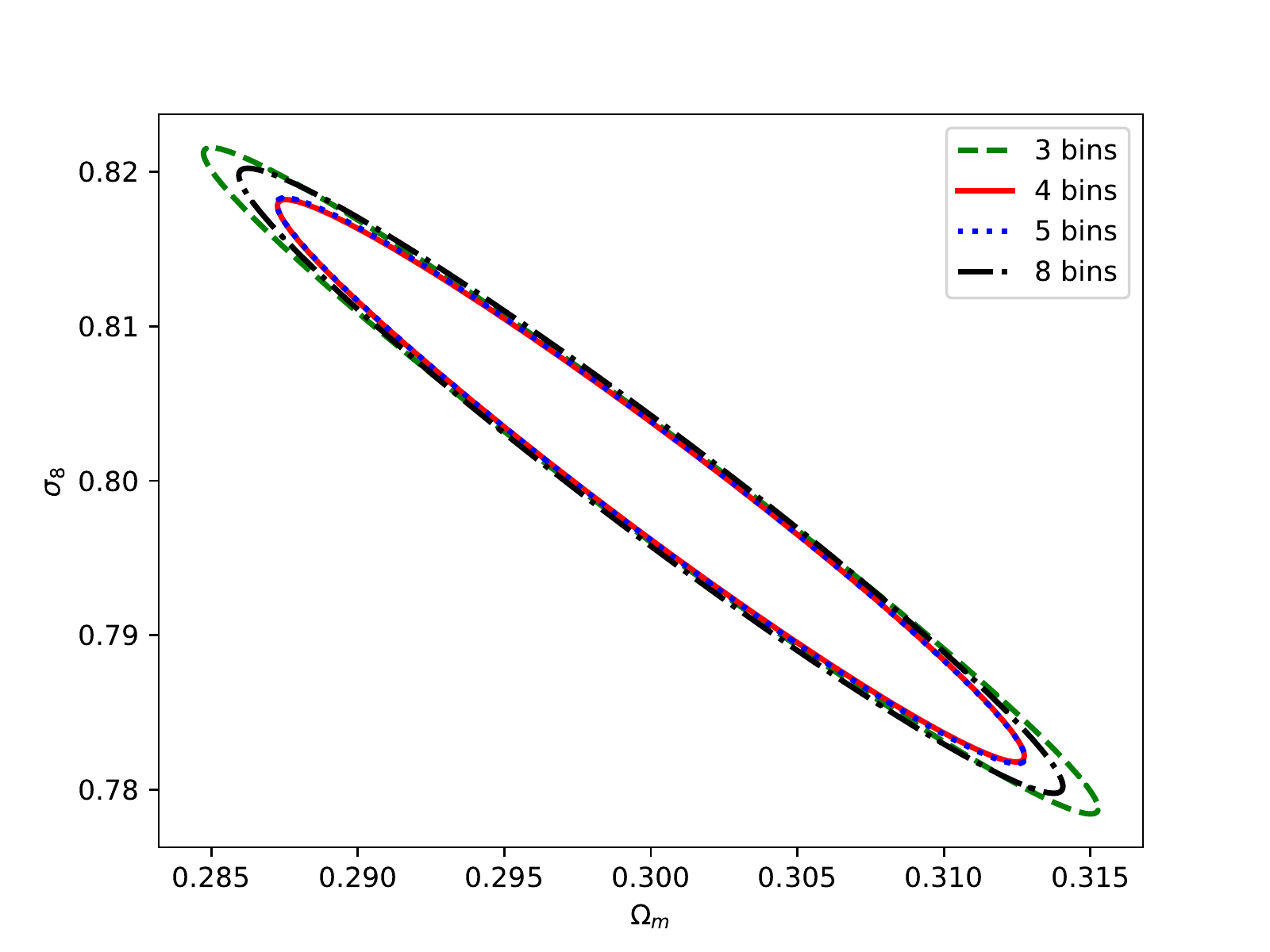}
\caption{Fisher forecast for different numbers of tomography bins. Photometric redshift error is taken to be $0.03(1+z)$.\label{fig:forecast}}
\end{center}
\end{figure}

We employ the Fisher forecast approach to compute the constraining power on the five dimensional parameter vector: $\mathbf{\theta} = (p_{\rm age}, \eta, h, \Omega_m, \sigma_8)$. The Fisher matrix is given by
\begin{equation}
  F_{IJ} = \frac{\partial \mathbf{d}^T}{\partial \theta_I} \mathrm{Cov}^{-1} \frac{\partial \mathbf{d}^T}{\partial \theta_J},
\end{equation}
where the data vector $\mathbf{d}$ is the collection of the observed power spectra $C_{\ell; i, j}^{\rm obs}$ and $\mathrm{Cov}$ is the covariance matrix given in Eq.~\eqref{eq:cov}. The covariance of parameter vector is estimated with the inverse of the Fisher matrix, $\mathrm{Cov}(\theta_I, \theta_J) \approx \left(F^{-1}\right)_{IJ}$.

We first study the dependence of the result on the number of redshift bins by comparing four binning schemes listed in Table~\ref{tab:bin}.
\begin{table}
  \begin{center}
    \caption{redshift binning schemes \label{tab:bin}}
  \begin{tabular}{ll}
    \hline
    \hline
    number of redshift bins & bin boundaries \\
    \hline
    $3$ & 0, 0.5, 1, 3 \\
    $4$ & 0, 0.5, 1, 1.5, 3 \\
    $5$ & 0, 0.4, 0.8, 1.2, 1.6, 3 \\
    $8$ & 0, 0.25, 0.5, 0.75, 1, 1.25, 1.5, 1.75, 3 \\
    \hline
  \end{tabular}
  \end{center}
\end{table}
The marginalized $68.3\%$ confidence-level constraints for $(p_{\rm age}, \eta)$, as well as the FROMAges for the four binning schemes are shown in the left panel of Figure~\ref{fig:forecast}. As we increase the number of redshift bins, the constraining power (FROMAge) increases at the beginning, and then drops when the photometric redshift error comes into play. A similar tendency is also observed for the other cosmological parameters, such as the $(\sigma_8, \Omega_m)$ combination presented in the right panel of Figure~\ref{fig:forecast}.

\begin{figure}
\begin{center}  
  \includegraphics[width=0.48\textwidth]{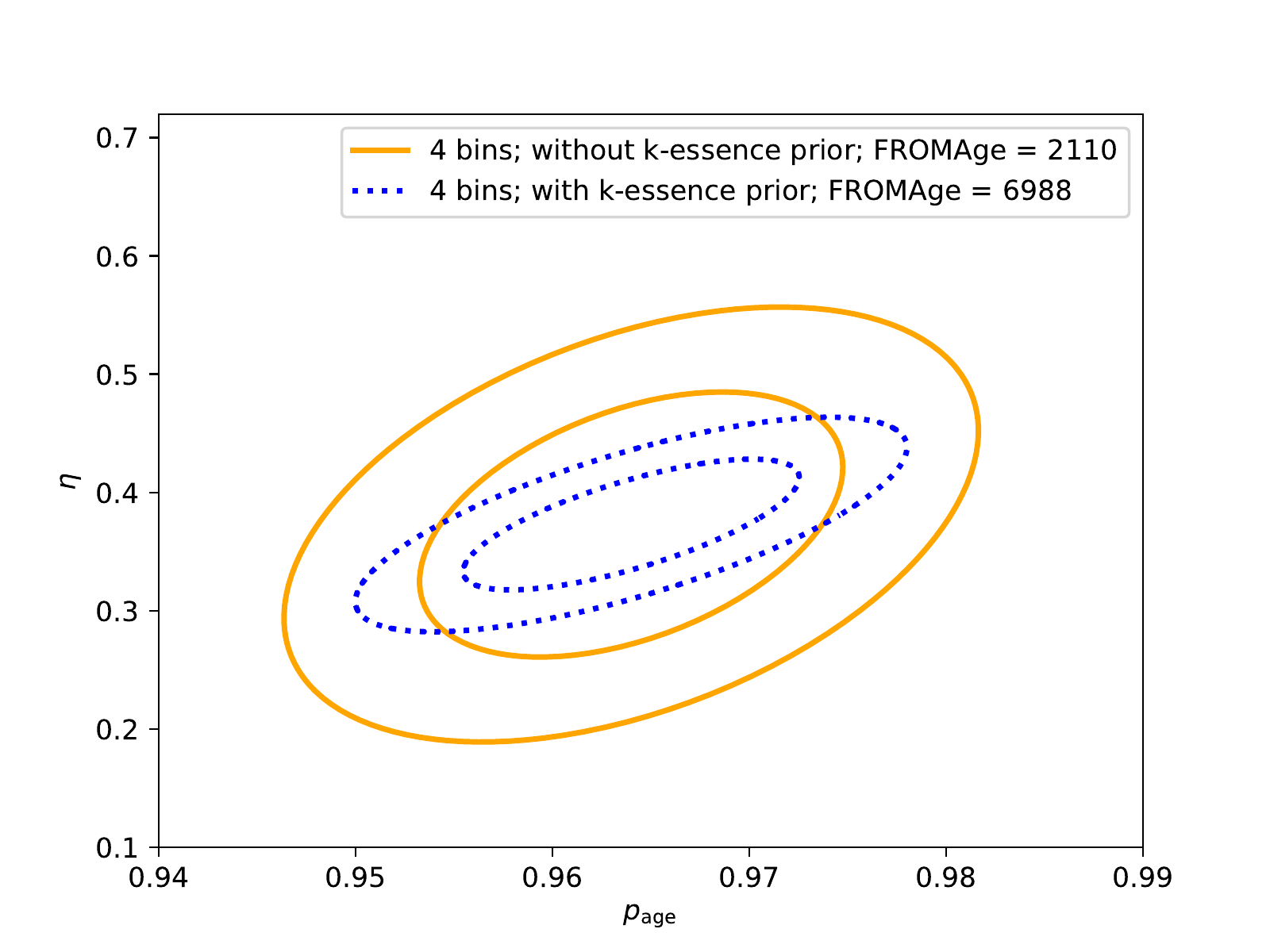}\includegraphics[width=0.48\textwidth]{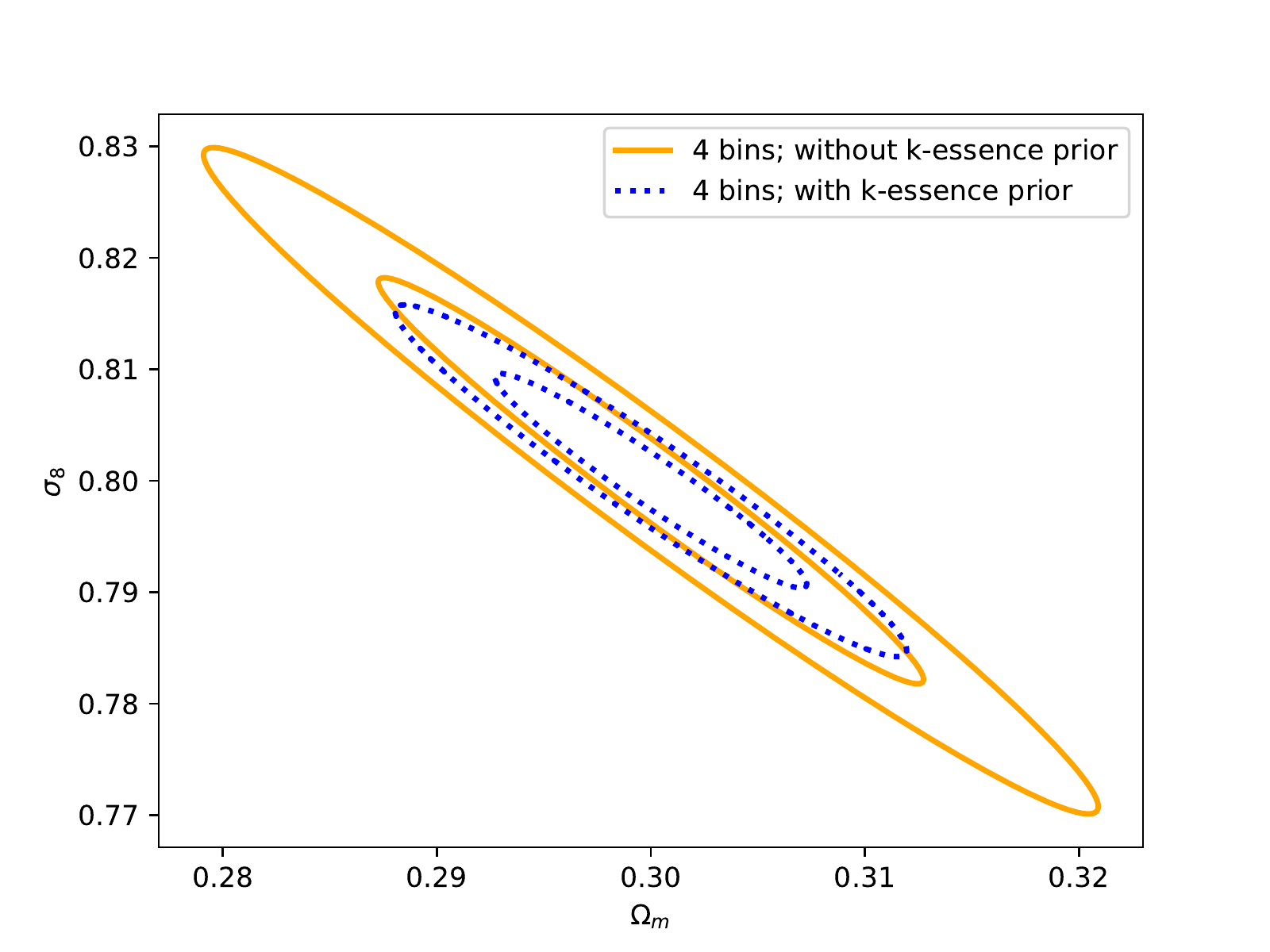}  
\caption{Fisher forecast of the $1\sigma$ and $2\sigma$ constraints on cosmological parameters, with and without thawing k-essence prior.\label{fig:prior}}
\end{center}
\end{figure}

Finally, we apply the thawing k-essence prior in the Fisher analysis. We first bin and interpolate a prior likelihood $P(\Omega_m, p_{\rm age}, \eta)$ from the random samples obtained in the previous section. A full likelihood is obtained by multiplying the data likelihood by the prior likelihood. We run Monte Carlo Markov Chain simulations to obtain the posterior covariance matrix, which is plotted in Figure~\ref{fig:prior} against the original Fisher forecast without thawing k-essence prior. For $(p_{\rm age}, \eta)$ the thawing k-essence prior improves the FROMAge by a factor of $3.3$. A similar improvement is found for $(\sigma_8, \Omega_m)$, too.

\section{Conclusions}

We have shown, with a simple Fisher forecast of future cosmic shear survey, that a reasonable theoretical prior of dark energy can significantly improve the constraining power of the data. This raises the question whether it is proper to judge the future dark energy surveys with a blind figure of merit without any theoretical prejudice. After all, the history of science has proven that theoretical prejudice is sometimes beneficial.

\section{Acknowledgements}

This work is supported by the National Natural Science Foundation of China (NSFC) under Grant No. 12073088, National SKA Program of China No. 2020SKA0110402,  National key R\&D Program of China (Grant No. 2020YFC2201600), and Guangdong Major Project of Basic and Applied Basic Research (Grant No. 2019B030302001).


\end{CJK*}
\end{document}